\documentclass[onecolumn, preprintnumbers, amsmath, amssym]{revtex4-1}

\usepackage{graphicx}% Include figure files
\usepackage{dcolumn}% Align table columns on decimal point
\usepackage{bm}% bold math

\usepackage{amssymb}
\usepackage{amsmath}
\usepackage{mathrsfs}
\usepackage{setspace}
\usepackage{multirow}
\begin{document}

\setstretch{1.5}

%\preprint{APS/123-QED}

\title{Rotation of pear-shaped $^{100}$Ru nucleus}% Force line breaks with \\

\author{A. Karmakar$^{1,2}$\thanks{\email anindita.karmakar@saha.ac.in},
P. Datta$^{3}$,\thanks,
Soumik Bhattacharya$^{2,4}$,
\altaffiliation[Present address: ]
{Department of Physics, Florida State University, Tallahassee, Florida, USA.}
Shabir Dar$^{2,4}$,
S. Bhattacharyya$^{2,4}$,
G. Mukherjee$^{2,4}$,
H. Pai$^{1,2}$,
\altaffiliation[Present address: ]
{Extreme Light Infrastructure - Nuclear Physics, Horia Hulubei National Institute for R$\&$D in Physics and Nuclear Engineering, Bucharest-Magurele, 077125, Romania.}
S. Basu$^{2,4}$,
S. Nandi$^{2,4}$,
\altaffiliation[Present address: ]
{Physics Division, Argonne National Laboratory, Lemont, Illinois 60439, USA.}
S. S. Nayak$^{2,4}$,
Sneha Das$^{2,4}$,
R. Raut$^5$,
S.S. Ghugre$^5$,
Sajad Ali$^{7}$,
R. Banik$^{8}$,
W. Shaikh$^{9}$,}

\author{S. Chattopadhyay$^{1,2}$}
\email{sukalyan.chattopadhyay@saha.ac.in}

\affiliation{$^{1}$Saha Institute of Nuclear Physics, 1/AF, Bidhan Nagar, Kolkata 700064, India}
\affiliation{$^{2}$Homi Bhabha National Institute, Training School Complex, Anushakti Nagar, Mumnai - 400094, India}
\affiliation{$^{3}$Ananda Mohan College, Kolkata- 700009, India.}
\affiliation{$^{4}$Variable Energy Cyclotron Centre, Kolkata - 700064, India}
\affiliation{$^{5}$UGC-DAE CSR, Kolkata Centre, Kolkata 700098, India}
\affiliation{$^{7}$Government General Degree College at Pedong, Kalimpong 734311, India}
\affiliation{$^{8}$Institute of Engineering $\&$ Management, Sector 5, Kolkata, India.}
\affiliation{$^{9}$Mugberia Gangadhar Mahavidyalaya, Purba Medinipur, India.}

\date{\today}% It is always \today, today,

\begin{abstract}%-------------------------------------------------------------------
Atomic nuclei in general can have deformed shapes and nearly all these shapes are symmetric
with respect to reflection. Only a few Actinide nuclei have stable reflection asymmetric pear shapes in their ground state and exhibit characteristic rotational bands. In this article, we report on the observation of two alternate parity rotational bands in $^{100}$Ru, which are connected by seven interleaved electric dipole transitions and their rates are found to be enhanced. In addition, the moments of inertia associated with these two opposite parity rotational bands have been found to be similar. These experimental observations indicate the rotation of a stable pear-shaped $^{100}$Ru nucleus, which is the first such observation outside the Actinide mass region. This shape is built on an excited configuration and originates from the rotational alignment of the angular momenta of a pair of neutrons. This unique observation establishes an alternate mechanism by which an atomic nucleus can assume a pear shape.
\end{abstract}

\maketitle

The loss of reflection symmetry is known to modify the properties of bulk materials as well as quantum systems. The polar acentric crystal classes break the point reflection symmetry. This symmetry states that if there is an atom at (x,y,z) relative to the centre of symmetry, there must also exist an atom at (-x,-y,-z) and this is true for all x,y,z. These polar crystals possess a dipole moment and exhibit technologically important properties like ferroelectricity and pyroelectricity \cite{coulson_1958}. In contrast, the reflection symmetry in a quantum system is broken due to the shape anisotropy. For example, the ZnO nano-prisms possess an electric dipole moment, which influences the band structure thereby modifying their optical properties \cite{article2}. The other well-studied quantum system of this class is the pear-shaped atomic nucleus, which also possesses an intrinsic dipole moment. It arises due to the separation between the centre of mass and the centre of charge as the concentration of protons is more in the region of higher curvature, which is the narrower end of the pear \cite{doi:10.1142/3530}. In recent years, these nuclei have attracted considerable experimental attention as the atoms with pear-shaped nuclei are ideal candidates for the search of permanent atomic electric dipole moment, which is indicative of CP violation and physics beyond the Standard Model \cite{RePEc:nat:nature:v:497:y:2013:i:7448:d:10.1038_nature12073, Chishti:2019emu, Butler2019}.\par
\begin{figure}[!ht]
    \begin{center}
       \hspace*{-0.2cm}\includegraphics[height=14cm, width=10cm, angle =0]{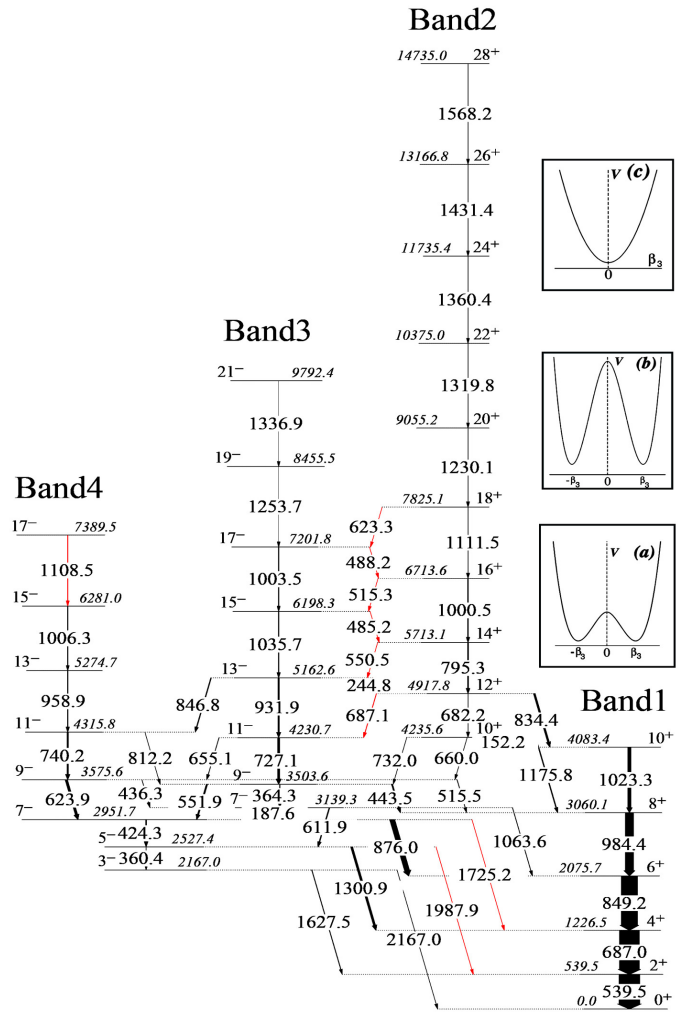}
    \end{center}
    \caption{The partial level scheme of $^{100}$Ru established from the present work. The ten gamma transitions marked in red, are new placements. The level and transition energies are expressed in keV and the uncertainties in the transition energies are $\pm$ 0.2 keV. The high spin levels beyond I = 22$\hbar$ were reported in Ref.~\cite{PhysRevC.62.044317} but not observed in the present experiment. A schematic representation of the evolution of the nuclear potential energy with increasing angular momentum is shown in the three insets.}
    \label{level_scheme}
\end{figure}

The pear shape of a nucleus can be realized by superimposing octupole deformation (characterized by $\beta_\textrm{3}$) on a  prolate shape (characterized by $\beta_\textrm{2}$) and its rotation for an even-even nucleus is characterized by a unique band structure, where the levels of two alternating parity bands are connected by relatively fast electric dipole (E1) transitions. Such a band structure was first reported in $^{218}$Ra \cite{FERNANDEZNIELLO1982221} and since then, the stable octupole deformation has been reported in a number of even-even isotopes of Ra – Th \cite{Chishti:2019emu, RePEc:nat:nature:v:497:y:2013:i:7448:d:10.1038_nature12073, PhysRevLett.78.2920, PhysRevLett.124.042503} (Z $\approx$ 88 and N $\approx$ 134). These nuclei possess the octupole deformation in their ground state due to the long-range octupole-octupole correlations among the nucleon orbitals close to the Fermi surface, whose total (J) and orbital (L) angular momenta differ by 3$\hbar$. The origin of the band structure of a rotating pear-shaped nucleus can be understood from the variation of the nuclear potential energy of this reflection-asymmetric shape as a function of the octupole deformation parameter. This is shown in the inset (a) of Fig.~\ref{level_scheme}. The potential energy has two degenerate minima at $\pm$$\beta_3^{min}$ separated by a finite barrier at $\beta_3$ = 0. The opposite parity bands correspond to the rotations of the two mirror shapes in these two minima. Thus, for stable octupole deformation, the moments of inertia (MOI) for the two opposite parity bands are expected to be identical. However, as the barrier height is finite, tunnelling is possible between the two minima, which leads to parity splitting in the laboratory frame. The review articles \cite{RevModPhys.68.349, Butler:2016rmu, article, Ahmad:1993nx} give a comprehensive account of the experimental and theoretical studies on the reflection asymmetric nuclei. \par

There exist other nucleon numbers namely, 34 and 56, where the octupole shape can be favoured \cite{RevModPhys.68.349, Butler:2016rmu, article, Ahmad:1993nx}. However, a recent theoretical study concluded that the nuclei of A $\approx$ 90 region are unlikely to possess stable octupole deformation \cite{PhysRevC.102.024311}. This is consistent with the observation that no well-defined rotational band structures have been found in the two N = 56 isotones namely, $^{96}$Zr (Z = 40) \cite{PhysRevC.42.R811} and $^{98}$Mo (Z = 42) \cite{PhysRevC.75.014314}, although they exhibit the existence of the octupole degree of freedom. On the other hand, the next isotone $^{100}$Ru (Z = 44), has two interspaced opposite parity bands beyond I = 11$\hbar$ and an E1 transition connecting the 12$^+$ and 11$^-$ levels was also reported \cite{PhysRevC.62.044317}. This led us to search for the inter-leaved E1 transitions between the levels of the opposite parity bands, whose presence would indicate a novel excited octupole band in $^{100}$Ru.\par

\begin{figure}[!ht]
    \begin{center}
       \hspace*{-0.3cm}\includegraphics[height=8cm, width=10cm, angle =0]{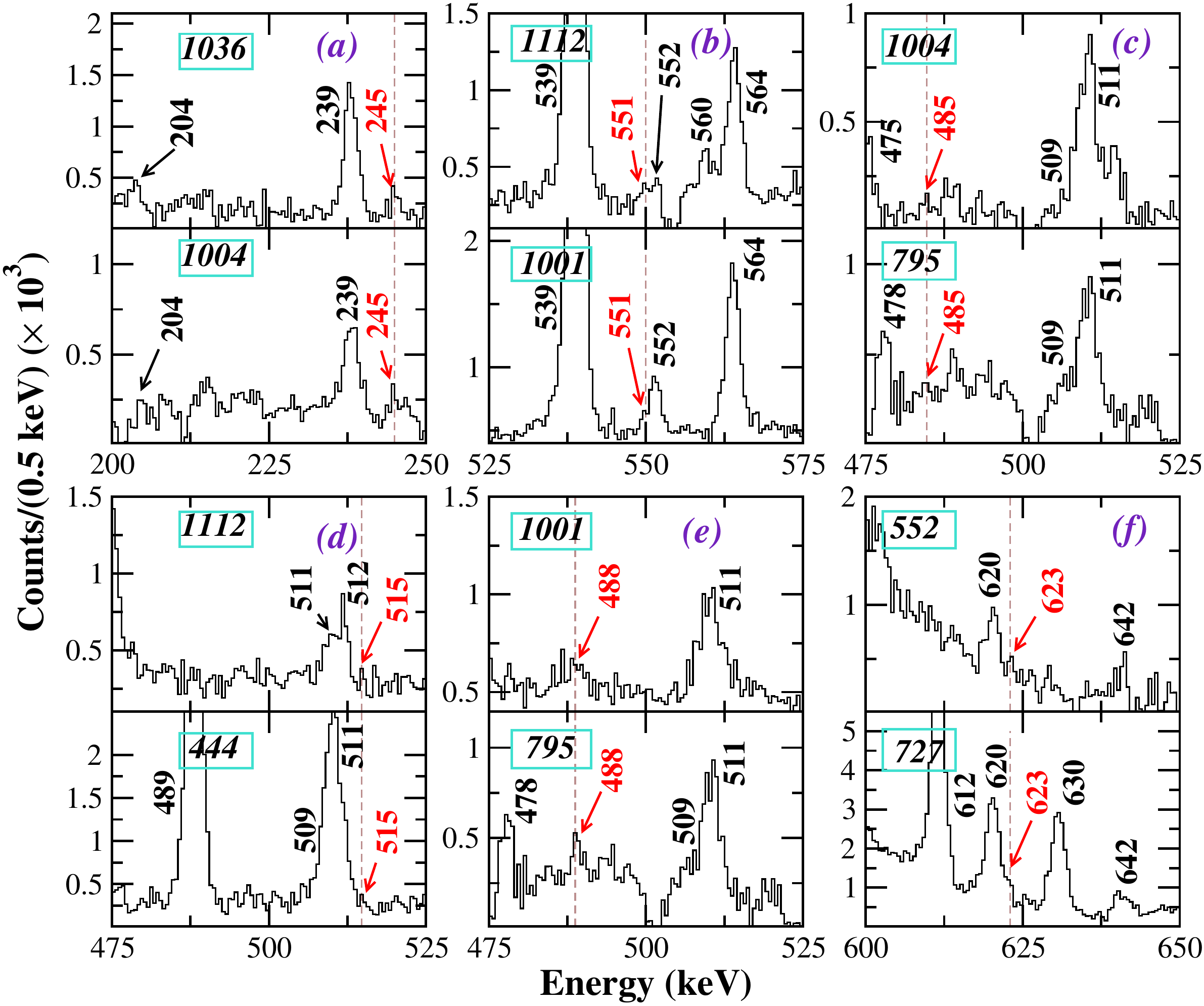}
    \end{center}
    \caption{The coincident gamma spectra observed with the detectors at the 90$^\circ$ ring. The E1 transitions connecting Band2 and Band3, are marked by the red dotted line. Each E1 transition is shown in two gamma-gated spectra. The numbers in the rectangular box represent the gamma-ray energy of the gating transitions. The gamma energies have been rounded off to their closest integer.}
    \label{gated_spec}
\end{figure}

The excited levels of $^{100}$Ru were populated through the fusion-evaporation reaction involving a 50 MeV alpha beam and a 98$\%$ pure 2mg/cm$^2$ thick $^{100}$Mo target. The alpha beam was delivered by K-130 cyclotron at VECC, India. The reaction and the beam energy were so chosen that the yield of $^{100}$Mo was nearly 90$\%$ of the total fusion cross-section. The $\gamma$ rays were detected by an array of 11 HPGe clover detectors, six of which were placed at 90$^\circ$ to the beam direction as the dipole emission probability is maximum at this angle. The other two and three detectors were placed in 40$^\circ$ and 125$^\circ$ rings, respectively. The data was collected by the PIXIE-16 digitizer-based data acquisition system \cite{DAS2018138} under these optimum conditions, which were crucial to identify the weak E1 transitions. The time-stamped data were sorted into $\gamma$-$\gamma$ matrices with a coincidence window of 200 ns using PIXSORT \cite{DAS2018138} and about 5 billion 2-fold $\gamma$-$\gamma$ coincidence events were recorded. The symmetric matrix was analysed using the RADWARE program ESCL8R \cite{RADFORD1995297} to build the partial level scheme of $^{100}$Ru, which is shown in Fig.\ref{level_scheme}. The thicknesses of the arrows are proportional to the relative intensities of the de-exciting gamma rays. The newly placed transitions from the present data are marked in red. The spins and parities of the excited levels determined from the present data \cite{supple} are consistent with those previously known \cite{PhysRevC.62.044317, GENILLOUD20003}. The intensity of each of the E1 transitions between Band2 and Band3 was determined from two gamma-gated spectra at 90$^\circ$. These gamma-gates are shown in Fig.~\ref{gated_spec}. The details of the analysis are given in Ref.~\cite{supple}. \par
%---------------------------------------------------------------------------------------------
%---------------------------------------------------------------------------------------------
The observation of three E3 transitions from the successive 3$^-$, 5$^-$ and 7$^-$ levels is a rare phenomenon and establishes the presence of octupole collectivity at low spins in $^{100}$Ru. In the case of octupole deformation, the static electric dipole moment of the pear shape will induce relatively fast E1 transitions. The estimated average retardation of 10$^5$ in the E1 transition rates (B(E1)) for the 3$^-$ level \cite{SINGH20211} is two orders of magnitude larger than that observed for the neighbouring $^{96}$Zr nucleus, which exhibits octupole instability \cite{PhysRevC.42.R811}. Thus, the 3$^-$ level seems to originate due to the collective octupole vibration around a reflection symmetric shape. Beyond I = 7 $\hbar$, the negative parity levels form two band structures namely, Band3 and Band4. The Band4 originates due to the single particle excitation (commonly known as ‘strongly-coupled band’), where one of the neutrons occupies the negative parity h$_{11/2}$ orbital. It may be noted that the signature partner (even spin negative parity band) of Band4 has been reported earlier \cite{PhysRevC.62.044317} and also established from the present data \cite{supple}. On the other hand, no signature partner could be identified for Band3, which indicates that it is a decoupled band and originates from a Rotational Aligned (RAL) configuration. The positive parity Band2 has also been proposed to originate from a RAL configuration of a pair of neutrons in the h$_{11/2}$ orbitals \cite{PhysRevC.62.044317}. The most significant feature of the present level scheme (Fig.~\ref{level_scheme}) is the observation of seven inter-leaved E1 transitions between Band2 and Band3. This observation indicates that these two opposite parity decoupled bands originate from a common mixed-parity RAL configuration. Such a configuration is possible due to the presence of octupole-driving orbitals of h$_{11/2}$ - d$_{5/2}$ states for N = 56 \cite{PhysRevC.45.2226}. It is well known that RAL can induce a modification of the quadrupole equilibrium deformation through polarization of the core \cite{Aberg:1990uh}. Thus, the RAL configuration for $^{100}$Ru may lead to a pear shape through the core polarization due to the presence of strong octupole correlations. A similar observation of two interspaced opposite parity bands connected by three E1 transitions from positive to negative parity levels has been reported after the neutron alignment in $^{66}$Zn \cite{ PhysRevC.104.044302, PhysRevC.102.064313}. It is interesting to note that in this case (N = 36), the octupole-driving orbitals of g$_{9/2}$ - p$_{3/2}$ states lie in close proximity. However, the low energy E1 transitions from negative to positive parity levels have not been observed. Thus, further experimental investigations are necessary to explore the origin of the enhancement octupole correlations in $^{66}$Zn after the rotational alignment.\par
%---------------------------------------------------------------------------------------------
%---------------------------------------------------------------------------------------------

\begin{figure}
    \begin{center}
       \hspace*{-0.5cm}\includegraphics[height=6cm, width=12cm, angle =0]{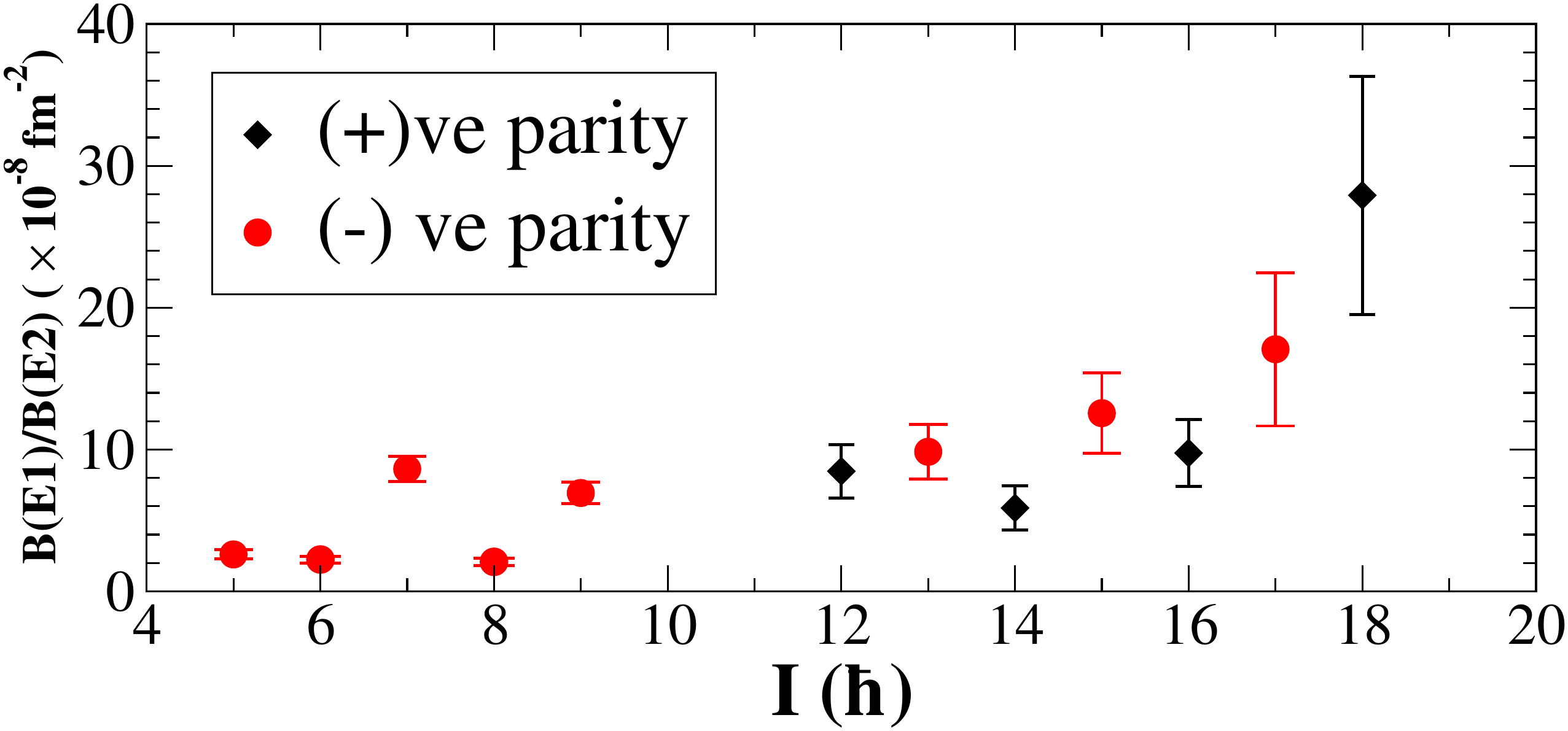}
    \end{center}
    \caption{The ratio of the inter-band E1 and the intra-band E2 transition rates from the excited levels of $^{100}$Ru as a function of spin, I.}
    \label{doqo}
\end{figure}

In order to explore the presence of octupole correlations in $^{100}$Ru, the ratio of the inter-band E1 and intra-band E2 transition rates (B(E1)/B(E2)) has been extracted from their measured intensities \cite{supple} and plotted in Fig.~\ref{doqo}. The B(E1)/B(E2) values for 6$^-$ and 8$^-$ levels have been extracted from the present data by assuming pure E1 character for the 6$^-$ $\xrightarrow{}$ 6$^+$ and 8$^-$ $\xrightarrow{}$ 8$^+$ transition. The de-exciting transitions from these levels are shown in Ref.~\cite{supple}, where the branching ratios have also been tabulated. It may be noted that the B(E1)/B(E2) values for the low spin levels are fairly constant except for the 7$^-$ level. If we assume Q$_0$ $\approx$ 200 efm$^2$ for $^{100}$Ru following the systematics of the B(E2, 2$^+$ $\xrightarrow{}$ 0$^+$) transition rate \cite{RAMAN19871}, the estimated average retardation of E1 rate is 10$^5$. This is the same as that estimated from the measured lifetime of 3$^-$ level \cite{SINGH20211}. Thus, the 5$^-$ level seems to originate also due to octupole vibration. On the other hand, the B(E1)/B(E2) values  for I $\geq$ 12$\hbar$ (weighted mean is 8.93 $\pm$ 0.87 fm$^{-2}$) are found to be significantly larger than those observed for low spin levels (weighted mean is 2.26 $\pm$ 0.15 fm$^{-2}$). This indicates the presence of octupole correlations in both Band2 and Band3. In this regard, it may be noted that the 10$^+_\mathrm{{Band2}}$ $\xrightarrow{}$ 9$^-_\mathrm{{Band3}}$ E1 transition (660.0 keV) has also been observed, but the non-yrast 8$^+$ level of Band2 could not be established from the present data. The estimated B(E1)/B(E2) value for the 7$^-$ level is higher than the other low spin levels and is similar to the observed values for Band2 and Band3. At the same time, this level also decays through an E3 transition (1725.2 keV), which is indicative of the octupole vibration. These observations probably indicate a strong mixing between the vibrational 7$^-$ level and the 7$^-$ level of Band3, which is also supported by the observation of the 187.6 keV transition between these two levels. In this case, there is an emergence of octupole correlation in $^{100}$Ru around I = 7$\hbar$. However, the alternating parity structure of Band2 and Band3 gets established only above I = 11$\hbar$ and the interleaved E1 transitions are observed beyond this spin. Thus, the observed band structure and the transition rates seem to suggest an evolution of the octupole degree of freedom in $^{100}$Ru – from vibrations at low spins (3$^-$ and 5$^-$ levels) to correlations in the intermediate (7$\hbar$ $\leq$ I $\leq$ 11$\hbar$) and deformation at higher spins (12$\hbar$ $\leq$ I $\leq$ 22$\hbar$).\par

\begin{figure}[!ht]
    \begin{center}
    \hspace*{-0.5cm}\includegraphics[height=9.0cm, width=13.0cm, angle =0]{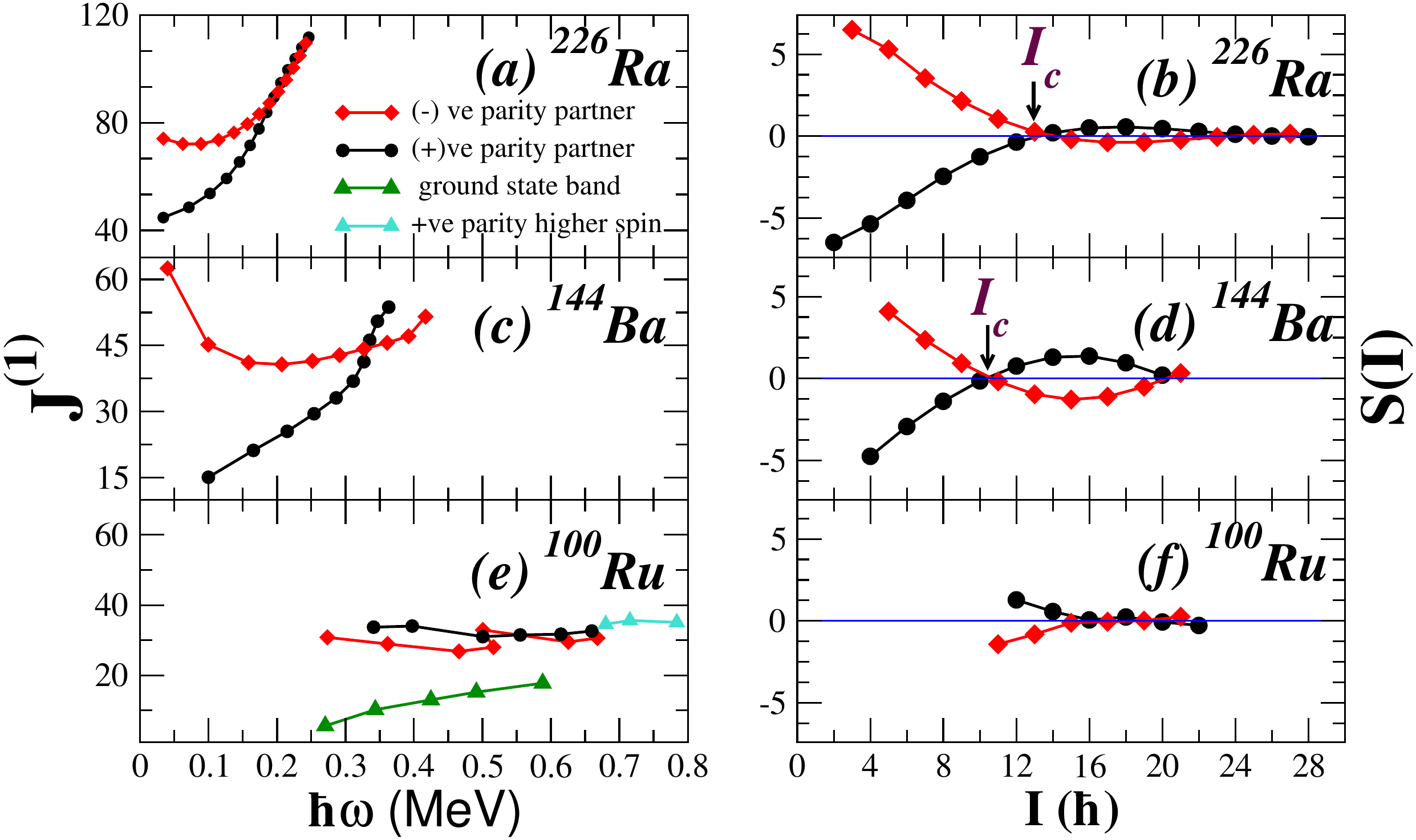}
    \end{center}
    \caption{The moment of inertia J(1) and the parity splitting index S(I) for the positive (black dots) and the negative parity (red diamonds) partner bands of $^{226}$Ra, $^{144}$Ba and $^{100}$Ru are plotted on the left and right panels as a function of rotational frequency ($\omega$) and angular momentum I, respectively.}
    \label{si}
\end{figure}

In order to explore the stability of the pear shape of $^{100}$Ru, the moments of inertia of the parity partner bands of the two well-known pear-shaped nuclei namely, $^{226}$Ra \cite{PhysRevLett.78.2920}, and $^{144}$Ba \cite{PhysRevLett.124.032501}, have been compared with those of $^{100}$Ru on the left panel of Fig.~\ref{si}. On the right panel, the values of parity splitting indices, S(I$^+$) and S(I$^-$) have been plotted, where S(I) is defined as the difference of the energy difference of the I, (I - 1) and (I – 2) levels \cite{PhysRevC.49.R605}.\par

For $^{226}$Ra, the value of both the parity indices becomes zero at I$_c$ = 12$\hbar$, which indicates the onset of strong octupole correlations \cite{PhysRevC.49.R605}. At the corresponding frequency ($\hbar\omega$ = 0.18 MeV), the MOI values of both the bands become similar as shown in Fig.~\ref{si}(a). This marks the onset of stable octupole deformation in $^{226}$Ra \cite{Ahmad:1993nx} and beyond  I$_c$, the negative parity levels become favoured in energy. This phenomenon of parity inversion in the nuclei of the Actinide region has been explored by Jolos et.al \cite{PhysRevC.72.064312} within the framework of the particle-rotor coupling model. These calculations indicate that the intrinsic configuration for the alternating parity bands changes from a fully paired configuration (K = 0, where K is the projection of I along the symmetry axis of the nucleus) before the parity inversion to one with rotationally aligned nucleons (K $\neq$ 0) after the inversion. Thus, the rotational alignment leads to the stabilization of octupole deformation as the pairing correlations become weaker \cite{NAZAREWICZ1985420}. With increasing spin, the potential barrier height at $\beta_3$ = 0 increases and the parity splitting due to the tunnelling between the two minima vanishes beyond the I = 23 $\hbar$ in $^{226}$Ra, as seen in Fig.~\ref{si}(b). The behaviour of  $^{144}$Ba (Fig~\ref{si}(d)) is similar to that of $^{226}$Ra at lower spins with I$_{c}$ = 10$\hbar$. However, beyond this point ($\hbar\omega$ = 0.29 MeV), the MOI values of the two partners continue to be significantly different till the highest observed frequencies as seen in Fig.~\ref{si}(c). This observation seems to indicate that  $^{144}$Ba does not exhibit the rotation of a stable octupole shape.\par
%---------------------------------------------------------------------------------------------
%---------------------------------------------------------------------------------------------
The MOI and S(I) plots for $^{100}$Ru are distinct from the above two cases as seen in Fig.~\ref{si}. It is observed from Fig.~\ref{si}(e) that the MOI values of the two parity bands are similar over the entire range of observed spins. For $\hbar\omega$ $\leq$ 0.5 MeV, the MOI values differ by a small amount due to the presence of parity splitting as observed from Fig.~\ref{si}(f). Thus, at lower spins, 12$\hbar$ $\leq$ I $\leq$ 16$\hbar$, the barrier height at $\beta_3$ = 0 is finite, which leads to the tunnelling between the two minima as depicted in the inset (a) of Fig.~\ref{level_scheme}. It may also be noted that the behaviours of the S(I) plot for $^{100}$Ru are similar to that for $^{224}$Ra after the parity inversion as in this case, the intrinsic configurations are similar (K $\neq$ 0). These observations indicate the presence of stable octupole deformation in $^{100}$Ru as for the RAL configuration, the pairing becomes substantially weaker. This is manifested in the MOI values of $^{100}$Ru, which are nearly independent of the rotational frequency, as seen from Fig.~\ref{si} (e). It may be noted that the MOI values of the pear-shaped odd-mass nuclei, for example, $^{223}$Th \cite{DAHLINGER1988337}
and $^{225}$Th \cite{HUGHES1990275} also exhibit a similar weak dependence on rotational frequency due to the presence of the odd nucleon, which lowers the pairing correlations due to Pauli blocking.\par
%---------------------------------------------------------------------------------------------
%---------------------------------------------------------------------------------------------
The sudden increase in the MOI value of Band3 around $\hbar\omega$ = 0.5 MeV (Fig.~\ref{si} (e)) indicates a change in the intrinsic configuration involving the rotational alignment of two more nucleons. The nature of this alignment can be determined from the observed rotational alignment in Band4 which occurs at the same frequency as seen in Fig.~\ref{level_scheme}. This alignment in Band4 is expected to be proton alignment as the neutron alignments are excluded due to the Pauli Blocking by the unpaired neutrons in (d$_{5/2}$/g$_{7/2}$) and h$_{11/2}$ orbitals. Thus, for $\hbar\omega$ $\geq$ 0.5 ( I $\geq$ $16\hbar$), the intrinsic configuration corresponds to two pairs of rotationally aligned neutrons and protons. The observation of the two interleaving transitions of 488.6 and 623.9 keV from the 17$^-$ and 18$^+$ levels, respectively, are crucial to establish that the levels beyond I = $16\hbar$ are also generated by the rotation of a pear-shape. Fig.~\ref{si} (a) shows that the MOI values of the two parity partner bands of $^{100}$Ru become identical beyond I = 16$\hbar$ and the parity splitting vanishes. This indicates that the tunnelling between the two minima stops beyond $\hbar\omega$ = 0.5 MeV, which is depicted by the large barrier at $\beta_3$ = 0 in the inset (b) of Fig.~\ref{level_scheme}\par

It is observed from Fig.~\ref{level_scheme} that the alternating parity structure is lost beyond I = 22$\hbar$ (corresponding to $\hbar\omega$ = 0.66 MeV) as Band3 terminates abruptly and a new band structure arises beyond this spin, whose MOI value (shown in cyan in Fig.~\ref{si} (e)) is $\approx$ 10$\%$ larger. This band may be associated with the rotation of a reflection-symmetric shape. A similar observation has also been reported in $^{222}$Th \cite{PhysRevLett.75.1050}, where a reflection-symmetric band has been predicted \cite{NAZAREWICZ1987437} to become yrast beyond I = 25$\hbar$. This proposed shape evolution in $^{100}$Ru has been represented by the vanishing of octupole deformation as shown in the inset (c) of Fig.~\ref{level_scheme}.\par

In summary, seven interleaved E1 transitions have been observed among the two alternate parity bands of $^{100}$Ru, whose moments of inertia are nearly identical. These observations are indicative of a stable octupole deformation, which originates due to the rotational alignment of a pair of neutrons in the presence of strong octupole correlations. In this case, the observed stability of the shape can be attributed to the substantial loss in the pairing correlation for the RAL configuration. These experimental observations indicate the presence of an alternate mechanism for the formation of a stable reflection asymmetric shape in an atomic nucleus.

\begin{acknowledgments}
The authors would like to thank the operational staff of the K-130 cyclotron at VECC for providing good quality beam as well as necessary support during the pandemic period. We are thankful to A. Navin for stimulating discussions. The authors are thankful to the Department of Atomic Energy and the Department of Science and Technology, Government of India for providing the necessary funding for the Clover array. A.K. acknowledges the grant from the Council of Scientific Research (CSIR) (File No: 09/489(0121)/2019-EMR-I), Government of India. 
\end{acknowledgments}

\bibliographystyle{apsrev4-1}

\begin{figure}[!ht]
    \begin{center}
       \hspace*{-0.5cm}\includegraphics[height=11.0cm, width=16cm, angle =0]{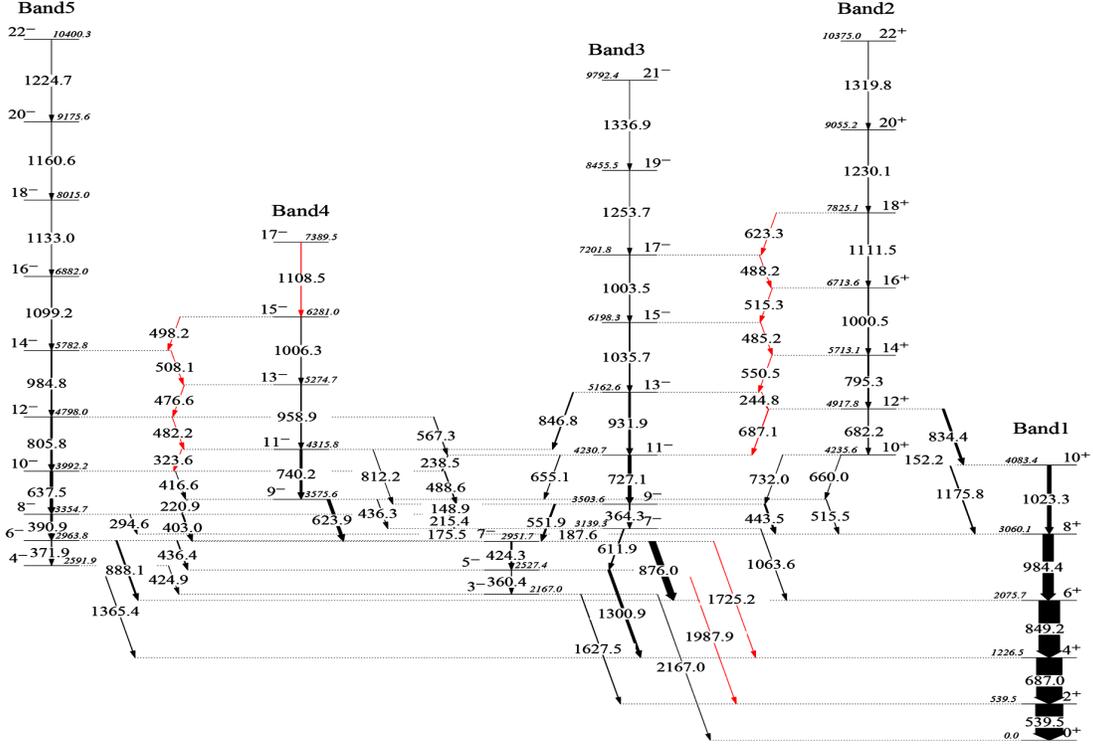}
    \end{center}
    \caption{The extended level scheme of $^{100}$Ru established from the present data}
    \label{level_ext}
\end{figure}

\newpage
\begin{center}
    \Large{\textbf{Rotation of pear-shaped $^{100}$Ru nucleus : Supplementary Information}}
\end{center}
The observed band structures of $^{100}$Ru in the present experiment are shown in Fig.~\ref{level_ext}. The gamma transitions marked in red, were assigned during the present work. The presence of M1 transitions between Band4 and Band5 up to I = 15$\rm{\hbar}$ establishes them as the signature partners.
Therefore, they originate from a strongly-coupled configuration. In this mass region, the negative parity signature partners arise from $\nu$h$_{11/2}$ $\bigotimes$ $\nu$(g$_{7/2}$/d$_{5/2}$) configuration. On the other hand, no signature partner for Band3 was reported and its search from the present data yielded null results.\par
%---------------------------------------------------------------------------------------------
%---------------------------------------------------------------------------------------------

\begin{figure}[!ht]
    \begin{center}
       \hspace*{-0.0cm}\includegraphics[height=7.5cm, width=17cm, angle =0]{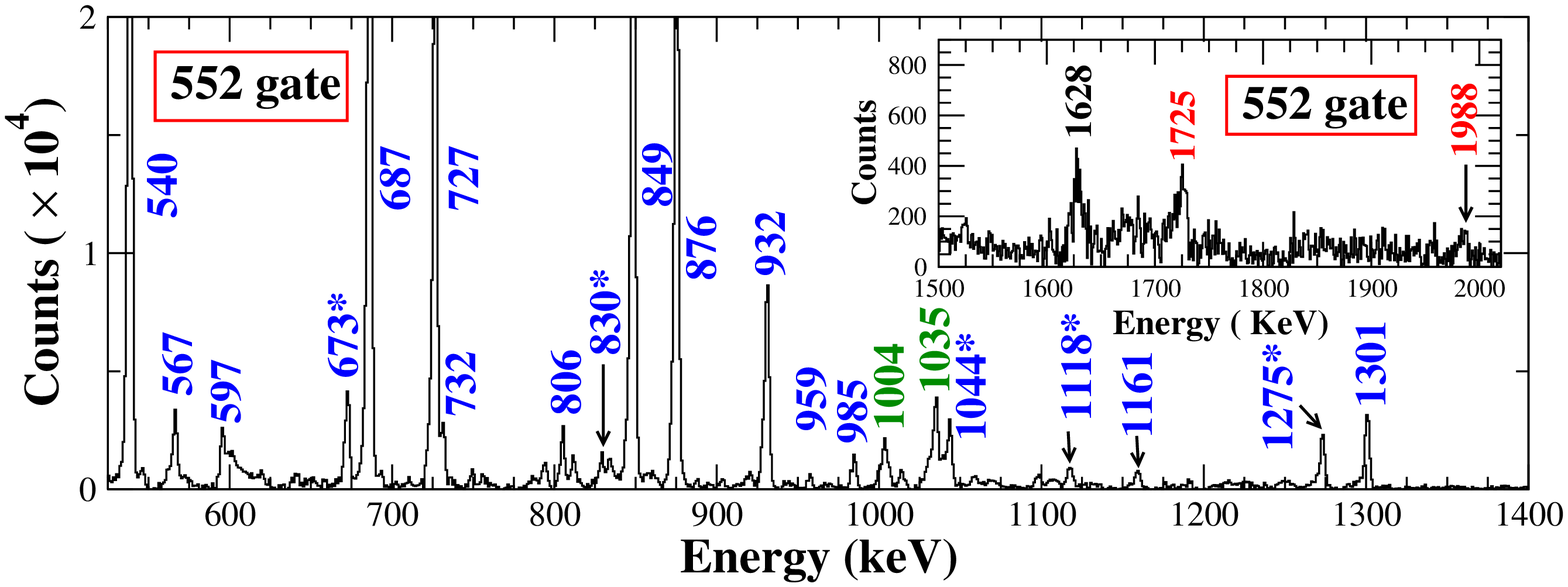}
     \end{center}
    \caption{The gamma-ray spectrum in coincidence with 552 keV (9$^-$ $\xrightarrow{}$ 7$^-$) transition. The gamma energies have been rounded off to the nearest integer.}
    \label{gate552}
\end{figure}

The two new E3 transitions observed in the present data are shown in the inset of Fig.~\ref{gate552}. It is observed from this figure that the 1035.7 keV transition is more intense than the 1003.5 keV transition. Thus, their placements were interchanged with respect to Ref.~\cite{PhysRevC.62.044317}. These two transition energies have been labelled in green. The gamma rays marked with * are from the excited levels of levels $^{100}$Ru, which are not shown in the partial level scheme of Fig.~\ref{level_ext}.\par

%---------------------------------------------------------------------------------------------
\begin{figure}[!ht]
    \begin{center}
       \hspace*{-0.0cm}\includegraphics[height=9.0cm, width=14cm, angle =0]{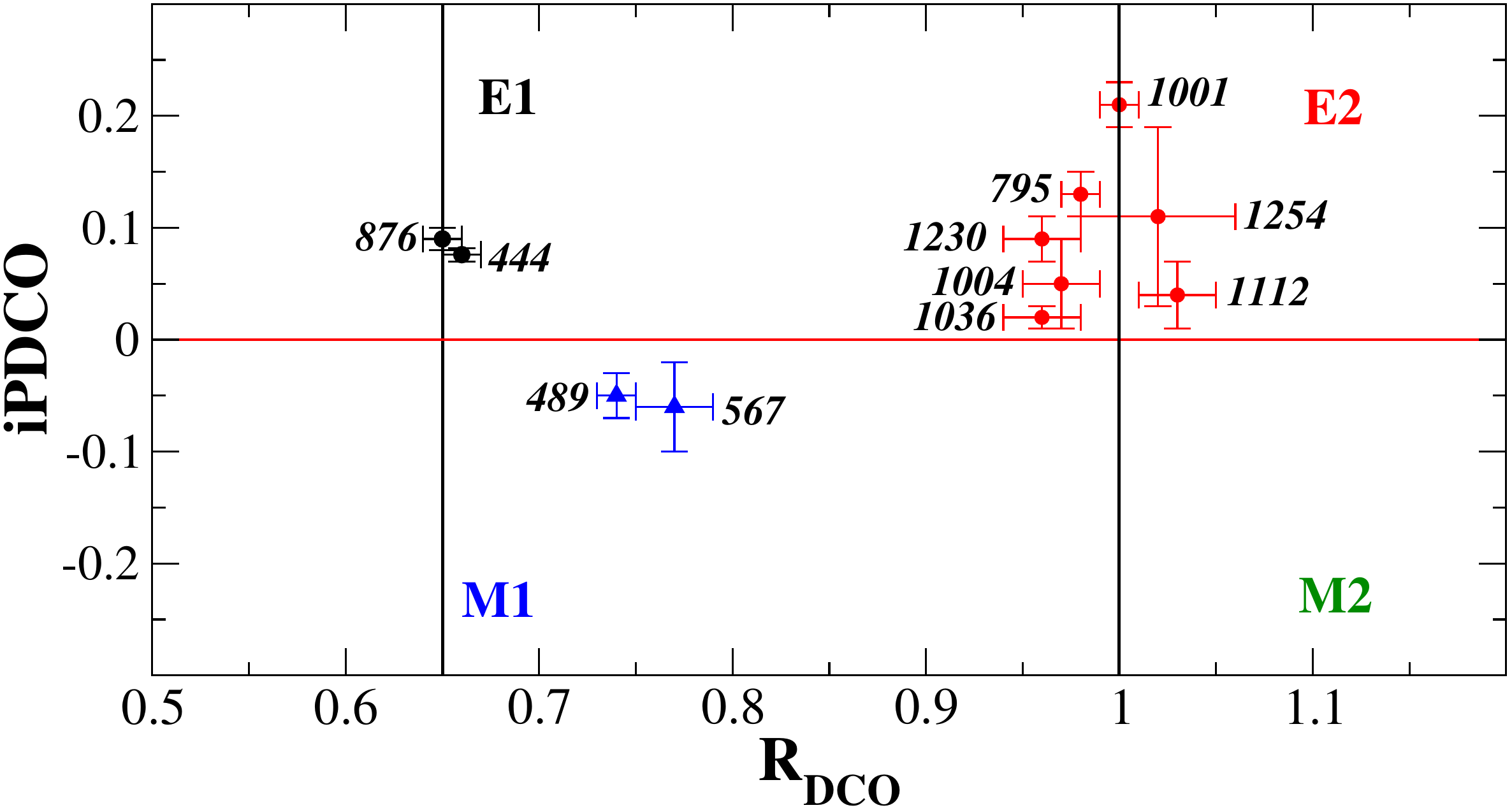}
     \end{center}
    \caption{The measured values of Ratio of Directional Correlations from Oriented states (R$\mathrm{_{DCO}}$) and integrated Polarization from the Directional Correlation of Oriented states (iPDCO) for different gamma transitions of $^{100}$Ru.}
    \label{dco}
\end{figure}

%---------------------------------------------------------------------------------------------
In order to measure the $\gamma$ -ray multipolarities by the Ratio of Directional
Correlations from Oriented states (R$_{DCO}$ ) method \cite{KRAMERFLECKEN1989333}, an angle-dependent matrix was constructed with the $\gamma$-ray energy detected at 90$^\circ$ along one axis while the coincident $\gamma$-ray energy at 55$^\circ$ on the other axis. For this combination of angles, the calculated R$_{DCO}$ values are 0.65 and 1.0 (indicated by the two vertical lines in Fig.~\ref{dco}) for pure $\Delta$I = 1 and $\Delta$I = 2 transitions, respectively, when a $\Delta$I = 2 gating transition is used. The linear polarization measurements were also performed to extract the electromagnetic character of the de-exciting $\gamma$ rays using the integrated Polarization from the Directional Correlation of Oriented states method (iPDCO) \cite{STAROSTA199916}. The analysis was performed using the present data sample for all the E2 transitions of the octupole band except for the 1336.9 keV transition of Band3, as the data were insufficient. The present measurements indicate the E1 character for 443.5 and 876.0 keV transitions, which establishes the negative parity assignment for Band3 of Fig.~\ref{level_ext}. The M1 character of 488.6 and 567.3 keV transitions (shown in Fig.~\ref{level_ext}) confirm the previous assignments \cite{GENILLOUD20003}. These results from the R$\mathrm{_{DCO}}$ and iPDCO measurements have been plotted in Fig.~\ref{dco}. These measurements are consistent with the previously known \cite{PhysRevC.62.044317, GENILLOUD20003} spin and parity assignments for the excited levels of $^{100}$Ru shown in Fig.~\ref{level_ext}. This establishes the E1 character of the inter-band transitions between Band2 and Band3 as the M2 mixing is expected to be negligible. It may be noted that the E2 character of the 1253.7 keV transition has been established from the present data set. The intensities of gamma rays in different gated spectra have been obtained by fitting the observed photo peaks to the Gaussian function using INGASORT software \cite{iaeaINISRepository}. The scheme for choosing of the $\gamma$-gates for the seven inter-leaved transitions is discussed below.\par
\textit{687.1 keV (12$^+$ $\xrightarrow{}$ 11$^+$ )}: All the top-gates show strong coincidence with the 687.0 keV transition (4$^+$ $\xrightarrow{}$ 2$^+$ ). Thus, the intensity has been estimated from the total feed-out intensity from the 11$^+$ level observed in
the 795 keV $\gamma$-gate. The 687.1 intensity was also estimated from the self-coincidence gate.\par
%---------------------------------------------------------------------------------------------
%---------------------------------------------------------------------------------------------

\textit{244.8 keV (13$^-$ $\xrightarrow{}$ 12$^+$ )}: The intensity has been estimated from the two top $\gamma$-gates of 1035.7 and 1003.5 keV. There is, however, contamination in these gates from another 245 keV transition, which feeds the 6$^+$ level with energy 2706 keV \cite{GENILLOUD20003}. This contamination of (37 $\pm$ 7)$\%$ was estimated from the observed feed-out intensities of the 6$^+$ level in the two gates.\par
\textit{550.5 keV (14$^+$ $\xrightarrow{}$ 13$^-$ )}: The intensity has been estimated from the two top $\gamma$-gates of 1000.5 and 1111.5 keV. This peak is well resolved from the 551.9 keV transition.\par
\textit{485.2 keV (15$^-$ $\xrightarrow{}$ 14$^+$ )}: The intensity has been estimated from the immediate top and bottom $\gamma$-gates of 1003.5 and 795.3 keV, respectively.\par
\textit{515.3 keV (14$^+$ $\xrightarrow{}$ 13$^-$ )}: The intensity has been estimated from the immediate top $\gamma$-gate of 1111.5 keV. In this gate, the contamination due to 515.5 keV transition ( 9$^-_\mathrm{{Band4}}$ $\xrightarrow{}$ 8$^+$ ) was found to be negligible. To validate this observation, the branching ratio for the 14$^+$ level was also evaluated in 443.5 keV gate, which is not in coincidence with the 515.5 keV transition.\par
\textit{488.2 keV (17$^-$ $\xrightarrow{}$ 16$^+$ )}: The intensity has been estimated from the two immediate bottom $\gamma$-gates of 1000.5 and 795.3 keV.\par
\textit{623.3 keV (18$^+$ $\xrightarrow{}$ 17$^-$ )}: The intensity has been estimated from the two bottom $\gamma$-gates of 551.5 and 727.1 keV since the higher gamma gates of Band3 are contaminated by the 623.9 keV transition ( 9$^-_\mathrm{{Band4}}$ $\xrightarrow{}$ 7$^-$ ). These two gating transitions are not in coincidence with the 623.9 keV transition.\par
The enhancement in the intensity of E1 transitions at 90$^\circ$ needs to be corrected for the estimation of the E1/E2 branching ratios. The correction factor was determined by measuring the branching ratios for (444.0 and 551.9 keV), (515.5 and 623.9 keV) and (876.0 and 424.3 keV) $\gamma$-rays at 90$^\circ$ and at 125$^\circ$ (at this angle the angular distribution effects are negligible). The weighted mean of the ratio of the branching ratios was found to be 1.58 $\pm$ 0.06. Thus, the E1/E2 branching ratios determined at 90$^\circ$ have been divided by this number to correct for the angular distribution effects. The E1 and E2 branching ratios for each level of the alternate parity bands were estimated from two different $\gamma$-gated spectra, which have been listed in Table.~\ref{table}. The B(E1)/B(E2) ratios for other low spin levels of $^{100}$Ru are listed in Table.~\ref{table2}.\par

The B(E1)/B(E2) rates were determined from the following relation.
\begin{equation}
\begin{aligned}
    \frac{B(E1, I_i \xrightarrow{} I_i-1)\downarrow}{B(E2, I_i \xrightarrow{} I_i-2)\downarrow} = \frac{1}{1.3\times10^6}\frac{I(E1)}{I(E2)}\frac{E_\gamma^5(E2)}{E_\gamma^3(E1)} fm^{-2}
    \label{be1}
\end{aligned}
\end{equation}
where, the energies of the $\gamma$-rays (E$_\gamma$) are expressed in MeV and I(E1)/I(E2) is the measured branching ratio.
\begin{table}[!ht]
\caption{The E1 and E2 branching ratios and the estimated values of B(E1)/B(E2) for the levels of the octupole band.}
\label{table}
\vspace{1mm}
\centering
\begin{tabular}{|c|c|c|c|c|}
\hline
E$_{level}$(keV) [J$^\pi$] & E$_\gamma$ (keV)  & $\gamma$-gates & I(E1)/I(E2) (10$^{-2}$) & B(E1)/B(E2) (10$^{-8}$ fm$^{-2}$) \\
\hline
{\multirow{2}{*}{4917.8 [12$^+$]}} & 687.1(E1)  & {\multirow{2}{*}{795.3 and 687.0}} & {\multirow{2}{*}{24.21 $\pm$ 5.37}} & {\multirow{2}{*}{8.47 $\pm$ 1.88}} \\
\cline{2-2}
& 682.2 (E2) &  &  &  \\
\hline
\hline
{\multirow{2}{*}{5162.6 [13$^-$]}} & 244.8 (E1)  & {\multirow{2}{*}{1035.7 and
1003.5}} & {\multirow{2}{*}{0.27 $\pm$ 0.05}} & {\multirow{2}{*}{9.84 $\pm$ 1.93}} \\
\cline{2-2}
& 931.9 (E2) &  &  &  \\
\hline
\hline
{\multirow{2}{*}{5713.1 [14$^+$]}} & 550.5 (E1)  & {\multirow{2}{*}{1111.5 and
1000.5}} & {\multirow{2}{*}{3.93 $\pm$ 1.04}} & {\multirow{2}{*}{5.88 $\pm$ 1.55}} \\
\cline{2-2}
& 795.3 (E2) &  &  &  \\
\hline
\hline
{\multirow{2}{*}{6198.3 [15$^-$]}} & 485.2 (E1)  & {\multirow{2}{*}{1003.5 and
795.3}} & {\multirow{2}{*}{1.04 $\pm$ 0.23}} & {\multirow{2}{*}{12.56 $\pm$ 2.83}} \\
\cline{2-2}
& 1035.7 ((E2) &  &  &  \\
\hline
\hline
{\multirow{2}{*}{6713.6 [16$^+$]}} & 515.3 (E1)  & {\multirow{2}{*}{1111.5 and 443.5}} & {\multirow{2}{*}{1.72 $\pm$ 0.41}} & {\multirow{2}{*}{9.75 $\pm$ 2.35}} \\
\cline{2-2}
& 1000.5 (E2) &  &  &  \\
\hline
\hline
{\multirow{2}{*}{7201.8 [17$^-$]}} & 488.2 (E1)  & {\multirow{2}{*}{1000.5 and
795.3}} & {\multirow{2}{*}{3.03 $\pm$ 0.98}} & {\multirow{2}{*}{17.07 $\pm$ 5.41}} \\
\cline{2-2}
& 1003.5 (E2) &  &  &  \\
\hline
\hline
{\multirow{2}{*}{7825.1 [18$^+$]}} & 623.3 (E1)  & {\multirow{2}{*}{727.1 and 551.9}} & {\multirow{2}{*}{5.17 $\pm$ 1.56}} & {\multirow{2}{*}{27.92 $\pm$ 8.40}} \\
\cline{2-2}
& 1111.5 (E2) &  &  &  \\
\hline
\end{tabular}
\end{table}

\begin{table}[!ht]
\caption{The E1 and E2 branching ratios and the estimated values of B(E1)/B(E2) for the low spin levels}
\label{table2}
\vspace{1mm}
\centering
\begin{tabular}{|c|c|c|c|c|}
\hline
E$_{level}$(keV) [J$^\pi$] & E$_\gamma$ (keV)  & $\gamma$-gates & I(E1)/I(E2) & B(E1)/B(E2) (10$^{-8}$ fm$^{-2}$) \\
\hline
{\multirow{2}{*}{2527.4 [5$^-$]}} & 1300.9 (E1)  & {\multirow{2}{*}{551.9}} & {\multirow{2}{*}{12.34 $\pm$ 1.51}} & {\multirow{2}{*}{2.61 $\pm$ 0.32}} \\
\cline{2-2}
& 360.4 (E2) &  &  &  \\
\hline
\hline
{\multirow{2}{*}{2963.8 [6$^-$]}} & 888.1 (E1)  & {\multirow{2}{*}{390.9}} & {\multirow{2}{*}{2.87 $\pm$ 0.30}} & {\multirow{2}{*}{2.25 $\pm$ 0.24}} \\
\cline{2-2}
& 371.9 (E2) &  &  &  \\
\hline
\hline
{\multirow{2}{*}{2951.7 [7$^-$]}} & 876.0 (E1)  & {\multirow{2}{*}{551.9}} & {\multirow{2}{*}{5.52 $\pm$ 0.57}} & {\multirow{2}{*}{8.63 $\pm$ 0.90}} \\
\cline{2-2}
& 424.3 (E2) &  &  &  \\
\hline
\hline
{\multirow{2}{*}{3354.7 [8$^-$]}} & 294.6 (E1)  & {\multirow{2}{*}{637.5}} & {\multirow{2}{*}{0.07 $\pm$ 0.01}} & {\multirow{2}{*}{2.08 $\pm$ 0.25}} \\
\cline{2-2}
& 390.9 ((E2) &  &  &  \\
\hline
\hline
{\multirow{2}{*}{3575.6 [9$^-$]}} & 443.5 (E1)  & {\multirow{2}{*}{727.1}} & {\multirow{2}{*}{1.24 $\pm$ 0.14}} & {\multirow{2}{*}{6.94 $\pm$ 0.76}} \\
\cline{2-2}
& 364.3 (E2) &  &  &  \\
\hline
\end{tabular}
\end{table}

\bibliographystyle{apsrev4-1}

\end{document}